\documentstyle[12pt,epsf]{article}

\def\gsim{\;\raise0.3ex\hbox{$>$\kern-0.75em\raise-1.1ex\hbox{$\sim$}}\;}
\def\lsim{\;\raise0.3ex\hbox{$<$\kern-0.75em\raise-1.1ex\hbox{$\sim$}}\;}
\newcommand{\be}{\begin{equation}}

\newcommand{\ee}{\end{equation}}
\newcommand{\bea}{\begin{eqnarray}}
\newcommand{\eea}{\end{eqnarray}}
\newcommand{\bt}{\begin{tabular}}
\newcommand{\et}{\end{tabular}}
\newcommand{\ba}{\begin{array}}
\newcommand{\ea}{\end{array}}
\newcommand{\ov}{\overline}

\newcommand{\bvec}{\mathbf}
\newcommand{\dps}{\displaystyle}

\begin{document}

\thispagestyle{empty}

\setcounter{page}{0}

{}\hfill{DSF$-$31/2003}

{}\hfill{physics/0401062}

\vspace{1truecm}

\begin{center}
{\Large \bf Following Weyl on Quantum Mechanics: the contribution
of Ettore Majorana}
\end{center}

\bigskip\bigskip

\begin{center}
{\bf A. Drago$^{1,a}$ and S. Esposito$^{1,2,3,b}$}

\vspace{.5cm}

$^!$ {\it Dipartimento di Scienze Fisiche, Universit\`{a} di
Napoli ``Federico II'' \\ Complesso Universitario di Monte S.
Angelo, Via Cinthia, I-80126 Napoli, Italy}

$^2$ {\it Istituto Nazionale di Fisica Nucleare, Sezione di
Napoli, Complesso Universitario di Monte S. Angelo, Via Cinthia,
I-80126 Napoli, Italy}

$^3$ {\it Unit\`a di Storia della Fisica, Facolt\`a di Ingegneria,
Universit\`a Statale di Bergamo, Viale Marconi 5, I-24044 Dalmine
(BG), Italy}

$^a$ e-mail address: adrago@na.infn.it

$^b$ e-mail address: Salvatore.Esposito@na.infn.it
\end{center}

\bigskip\bigskip\bigskip

\vspace{3cm}
\begin{abstract}
\noindent After a quick historical account of the introduction of
the group-theoretical description of Quantum Mechanics in terms of
symmetries, as proposed by Weyl, we examine some unpublished
papers by Ettore Majorana. Remarkable results achieved by him in
frontier research topics as well as in physics teaching point out
that the Italian physicist can be well considered as a follower of
Weyl in his reformulation of Quantum Mechanics.
\end{abstract}

${}$ \\


\newpage

\section{INTRODUCTION}

The important role of symmetries (and, then, of Group Theory) in
Quantum Mechanics was established in the third decade of the XX
century, when it was discovered the special relationships
concerning systems of identical particles, reflection and
rotational symmetry or translation invariance. In particular it
was realized that a given symmetry of (the Hamiltonian of) a
physical system leads to the partition of its states into terms of
different symmetrical behavior of their eigenfunctions and then to
selection rules. The systematic theory of symmetry resulted to be
just a part of the mathematical theory of groups (see, for
example, \cite{Weyl28}).

In many cases the dependence of the eigenfunctions on the
variables involved in the symmetry is explicit, and thus the
partitions into systems of different symmetry is straightforward.
However, this method fails if the number of variables is large. An
alternative approach is to characterize the behavior of the
eigenfunctions by means of transformation matrices; for example,
for a rotation along a given axis of an angle $\alpha$, the
corresponding matrix is: \be \left( \begin{array}{rr} \cos \alpha
& - \sin \alpha \\ \sin \alpha & \cos \alpha \end{array} \right) \
. \ee Matrices of this kind corresponding to a given group of
transformations form the ``representation of the group by means of
linear transformations'', and the sets of terms in the partitions
correspond to the ``irreducible'' representations of the group of
the covering operations of the given physical system. This form of
Group Theory was first applied to the quantum theory by E. Wigner
in 1926-1927 \cite{Wigner27}. The inclusion of the spin (and other
internal symmetries) into the game was, then, possible when H.
Weyl discovered the two-valued representations of the rotation
group \cite{Weyl28} and used them to describe the atomic states
with spin.

The final form of such applications of Group Theory to Quantum
Mechanics was established in the books by Weyl \cite{Weyl28},
Wigner \cite{Wigner31} and van der Waerden \cite{Waerden32}.
Nevertheless, despite these successes, the group-theoretical
description of Quantum Mechanics in terms of symmetries was
ignored by almost all theoretical physicists and it received
attention by physics textbooks only in recent times. It is also
noticeable that, although in the 1950's almost every physicist had
a copy of the Weyl book \cite{Yang}, the extensive use of Group
Theory in physics research started only in nuclear and particle
physics in 1950's.

Such an indifference, however, does not apply to Ettore Majorana.
Although he was in Germany for a short period \cite{unpub}, he
never met Weyl, but the brilliant works on Group Theory and its
applications of this mathematician, as well as those of Wigner,
left an unambiguous and fruitful mark on the work of the Italian
physicist. This is not easily recognizable by looking at the very
few published papers by Majorana (just 9 different articles, one
of which posthumous) but clearly emerges from his unpublished
manuscripts (see also the comment of N. Cabibbo in
\cite{lezioni}), most of which are deposited at the {\it Domus
Galilaeana} in Pisa (Italy) \cite{unpub}.

Majorana was introduced to Physics at the beginning of 1928 by E.
Segr\'e, and graduated with Fermi on July 6, 1929. He thus went on
to collaborate with the famous group in Rome created by Enrico
Fermi and Franco Rasetti. During 1933 Majorana spent about six
months in Leipzig with W. Heisenberg, and then, for some unknown
reasons, stopped participating in the activities of Fermi's group.
He even ceased publishing the results of his research, except for
his paper on the ``symmetric theory of electrons and positrons"
\cite{elpos}, which (ready since 1933) Majorana was persuaded by
his colleagues to remove from a drawer and publish just prior to
the 1937 Italian national competition for full-professorships.
With respect to the last point, it is remarkable that, on the
recommendation of the judging committee (Fermi being one of the
members), the Italian Minister of National Education installed
Majorana as professor of Theoretical Physics at the University of
Naples because of his ``great and well-deserved fame",
independently of the competition itself \cite{unpub}. Fermi was,
probably, the only great physicist of that time that adequately
recognized the genius of Majorana, as can be deduced from his own
words: ``Because, you see, in the world there are various
categories of scientists: people of a secondary or tertiary
standing, who do their best but do not go very far. There are also
those of high standing, who come to discoveries of great
importance, fundamental for the development of science. But then
there are geniuses like Galileo and Newton.  Well, Ettore was one
of them. Majorana had what no one else in the world had..."
\cite{fermicoc}. Unfortunately, Ettore Majorana disappeared during
rather mysteriously on March 26, 1938, and was never seen again.

We have analyzed the physics work by Majorana before and after the
appearance of the Weyl's book, and the results of this research
are presented here. We find that Majorana can be well considered
as a faithful follower (and, probably, the only one) of the Weyl
thought who, however, went beyond the track drawn by Weyl himself.
In the next section we outline the Weyl proposal for an adequate
description of Quantum Mechanics in terms of Group Theory and
discuss the results reached by the author himself. Instead in
Sect. \ref{3} we consider the work of Majorana in the direction
tracked by Weyl, while in the subsequent section we compare the
two approaches and the corresponding results obtained. Finally in
Sec. \ref{5} we give our conclusions.

${}$ \\

\section{\label{2} WEYL PROPOSAL}

According to Wigner \cite{Wigner31}, {\it it was probably M. von
Laue who first recognized the significance of Group Theory as the
natural tool with which to obtain a first orientation in problems
of Quantum Mechanics}. However, the program for a description of
Quantum Mechanics in terms of Group Theory is clearly stated for
the first time by Weyl in the introduction of his well-known book
\cite{Weyl28}:
\begin{quote}
``...it has recently been recognized that Group Theory is of
fundamental importance for quantum physics; it here reveals the
essential features which are not contingent on a special form of
the dynamical laws nor on special assumptions concerning the
forces involved. We may well expect that it is just this part of
quantum physics which is most certain of a lasting place. Two
groups, {\it the group of rotations in 3-dimensional space} and
{\it the permutation group}, play here the principal role... The
investigation of groups first becomes a connected and complete
theory in {\it the theory of the representations of groups by
linear transformations}, and it is exactly this mathematically
most important part which is necessary for an adequate description
of the quantum mechanical relations...''
\end{quote}
As an illustration, let us consider the problem of the splitting
of the spectral lines of an atom, placed in a homogeneous magnetic
field $B$ in the direction of the $z$ axis: for simplicity we
neglect the spin interaction with the magnetic field (Zeeman
effect). The non group-theoretical description of the phenomenon
proceeds as follows. Let us consider, at first, one-electron
atoms; the Hamiltonian describing the system is then:
\begin{equation}\label{2.1}
  H \; = \; H_0  \; + \; W \; = \; H_0 \; + \; \mu_B B L_z
  ,
\end{equation}
where $H_0$ is the undisturbed Hamiltonian of the electron and the
second term $W$ accounts for the ``perturbation'' introduced by
the magnetic field. Here $\mu_B = e \hbar / 2 m c$ is the Bohr
magneton for the electron and $L_z$ is the $z$ component of the
angular momentum operator divided by $\hbar$. The energy
eigenfunctions of the system coincide with the eigenstates
$\psi_m$ of the operator $L_z$, whose eigenvalues are the integers
$m$. Thus the energy terms results to be $E= E_0 + \mu_B B m$, and
the frequencies $\omega$ of the spectral lines corresponding to
the transition $E \rightarrow E^\prime$ are given by:
\begin{equation}\label{2.2}
  \hbar \omega \; = \; E \; - \; E^\prime \; = \; (E_0 \; - \;
  E^\prime_0) \; + \; \mu_B B (m - m^\prime) .
\end{equation}
Each spectral line is, then, broken up into the lines associated
with all possible transitions $m \rightarrow m^\prime$. Since $m$
may assume the $2 \ell +1$ values between $- \ell$ and $+ \ell$,
the integer $\ell$ being the orbital quantum number associated to
the eigenvalues of the $L^2$ angular momentum operator, in general
we can expect a splitting into $(2 \ell +1)(2 \ell^\prime +1)$
lines. Nevertheless the transition probabilities are proportional
to the squared modulus of the matrix elements of the components
$q_x, q_y, q_z$ of the dipole moment of the atom along the
coordinate axes, given by:
\begin{eqnarray}
  \left( q_x + i q_y \right)_{m^\prime m} &=& \int
  \psi^\ast_{m^\prime}\left( q_x + i q_y \right) \psi_m \ d^3
  \vec{r} , \label{2.3} \\
  \left( q_x - i q_y \right)_{m^\prime m} &=& \int
  \psi^\ast_{m^\prime}\left( q_x - i q_y \right) \psi_m \ d^3
  \vec{r} , \label{2.4} \\
  \left( q_z \right)_{m^\prime m} &=& \int
  \psi^\ast_{m^\prime}  q_z \psi_m \ d^3 \vec{r} . \label{2.5}
\end{eqnarray}
By expressing the integrand functions in polar coordinates $(r,
\theta, \phi)$, we see that the one in $\left( q_x + i q_y
\right)_{m^\prime m}$ contains the factor $\exp \left\{ i (-
m^\prime +1+m) \phi \right\}$ which, integrated with respect to
$\phi$, gives zero unless $m^\prime = m+1$. Similarly, for the
second and third integrand we have $m^\prime = m-1$ and $m^\prime
= m$, respectively. We thus find a splitting into only three
lines, corresponding to the selection rules $\Delta m = m^\prime -
m = 0, \pm 1$, as observed experimentally in the normal Zeeman
effect.

The group-theoretical description of the same phenomenon does not
need the previous simplification of one-electron atoms, but
considers the system as a whole without resolving it into
individual electrons. This is justified by the Wigner observation
in 1927 that it is always possible to choose the eigenfunctions
$\psi_m$ of the unperturbed Hamiltonian $H_0$ in such a way that a
rotation over an angle $\phi$ about the $z$ axis transforms
$\psi_m$ into $\exp \left\{- i m \phi \right\} \psi_m$, with
integer $m$. Given the rotational invariance and that $L_z$ is the
infinitesimal generator of rotations about the $z$ axis, we then
have, once more, $L_z \psi_m = m \psi_m$, so that Eq. (\ref{2.2})
again follows from Eq. (\ref{2.1}). As far as the transition
probabilities are concerned, we note that the component $q_z$ has
to remain unchanged when a rotation over an angle $\phi$ about the
$z$ axis is performed, so that $\left( q_z \right)_{m^\prime m}$
must be diagonal. On the other hand, the quantities $q_x \pm i
q_y$ acquire a factor $\exp \left\{ \mp i \phi \right\}$ when a
rotation is performed, while $\psi^\ast_{m^\prime}$ and $\psi_m$
account for a global factor of  $\exp \left\{i (m^\prime - m) \phi
\right\}$. From rotational invariance, then, the selection rules
immediately follow:
\begin{eqnarray}
  \left( q_x + i q_y \right)_{m^\prime m} : & \quad & \quad
  \Delta m \, = \, +1 , \label{2.6} \\
  \left( q_x - i q_y \right)_{m^\prime m} : & \quad & \quad
  \Delta m \, = \, -1 ,  \label{2.7} \\
  \left( q_z \right)_{m^\prime m} : & \quad & \quad
  \Delta m \, = \, 0 . \label{2.8}
\end{eqnarray}
When the spin interaction with the magnetic field is taken into
account, the anomalous Zeeman effect arises, whose explanation
requires to change the perturbation term $W$ in the Hamiltonian
(\ref{2.1}) and the formalism of relativistic Quantum Mechanics
should be used. Nevertheless the selection rules for the quantum
number $m$ established above have been obtained from fundamental
concepts of Group Theory, so that they are valid in all cases of
the Zeeman effect. It is then possible to deal with the splitting
of the spectral lines occurring when the symmetry is decreased; in
the case considered above, we have the transition from the
spherical symmetry of the undisturbed $H_0$ to the axial symmetry
of the perturbed $H$. In particular, Weyl recognized (see, for
example, Sect. A of Chapter IV in \cite{Weyl28}) that the breaking
up of the energy levels, due to the axially symmetric
perturbation, parallels the reduction of irreducible
representations of the rotation group, when this is restricted to
the group of rotations about the $z$ axis. The problem considered
above is, then, a typical example which well illustrates the
above-quoted Weyl proposal.

The Weyl book \cite{Weyl28} is practically a sandwich of
mathematical formalism and physics applications (starting from
mathematics); nowadays it can be considered quite a successful
attempt for an ``adequate'' group-based description of the quantum
mechanical phenomena known at that time. Nevertheless it was
readily recognized a great reluctance among physicists toward
accepting the group-theoretical point of view, they called it the
``group pest'' (see, for example, the prefaces of Refs.
\cite{Wigner31} and \cite{Weyl28} in their English translations).
Remarkably, in the second edition of his book, Weyl did not go
further in his program but ``followed the trend of the times, as
far as justifiable, in presenting the group-theoretic portions in
as elementary a form as possible''. However the word
``elementary'' used by Weyl is of dubious meaning; indeed, the
(assumed) didactic sense of this sentence is contradicted by a
careful look inside the revised book (for another interpretation
of the Weyl thought, in terms of his elementary mathematics, see
\cite{Drago}). After the publication of the second edition of his
book, Weyl abandoned his proposal in order to improve the
formulation of Group Theory and render it a more suitable tool for
Theoretical Physics; the ``group pest'' was, however, cut out from
fundamental physics for many years, until the discovery of
peculiar symmetries in nuclear and particle physics. Furthermore
we must wait up to recent times for group-based descriptions of
Quantum Mechanics in physics textbooks.

Probably the failure of the Weyl program should be looked for in
the involved mathematical presentation of the subject, which
seemed too much hard to swallow to the physicists of that time
\cite{schrodinger}. In this depressing framework it appears even
more relevant the enthusiasm by Majorana for Weyl's proposal about
group-theoretical foundation of Quantum Mechanics.

${}$ \\

\section{\label{3} MAJORANA ON GROUP THEORY AND ITS APPLICATIONS}

Among the very few books owned by E. Majorana (about 15 volumes
kept by his nephew Ettore in Rome) it appears the Weyl one
``Gruppentheorie und Quantenmechanik'' in its first German edition
(1928). As testified by the large number of unpublished manuscript
pages (and also some published articles) of the Italian physicist,
the Weyl approach greatly influenced the scientific thought and
work of Majorana. Here we focus primarily on his 5 orderly
notebooks, known as the ``Volumetti'' \cite{volumetti}, written
between 1927 and 1932. Besides a surprising clarity and linearity
of exposition, two peculiar features of the ``Volumetti" are
relevant for us: they are dated by the author himself (each
notebook was written during a period of time of about one year,
starting from 1927) and explicit references to the Weyl book are
included \footnote{In about 500 pages, only 5 bibliographical
references appears, 3 of which correspond to the Weyl book of
1928.}. These features allow us to reconstruct the line of
thinking of our author. In fact, just by looking at the table of
contents of the notebooks (this table was written down by Majorana
himself) we are able to realize the impact of Weyl's work.

The first two notebooks, whose ``closing dates'' correspond to
March 1927 and April 1928, respectively, deal mainly with
electromagnetism, thermodynamics and atomic physics arguments,
accounting for a total of 81 sections, and no reference to Group
Theory or some related application occurs. The situation changes
starting from the third notebook, whose closing date is June 1929:
a number of sections are devoted to the crucial topic under
consideration.

In the following we report the list of such sections pertaining to
the last three notebooks (out of a total of 64 sections):
\\ ${}$ \\
VOLUMETTO III
\begin{itemize}
 { \item[{13.}]The group of proper unitary transformations in two variables }{}
 { \item[{14.}]Exchange relations for infinitesimal transformations in the representations of continuous groups }{}
 { \item[{16.}]The Group of Rotations $O(3)$ }{}
 { \item[{17.}]The Lorentz Group }{}
 { \item[{18.}]Dirac matrices and the Lorentz Group }{}
 { \item[{20.}]Characters of ${ D}_j$ and reduction of ${ D}_j {\times } { D}_j^{\prime }$ }{}
 { \item[{21.}]Intensity and selection rules for a central field }{}
 { \item[{22.}]The anomalous Zeeman effect (according to the Dirac theory)}{}
 \\
 \end{itemize}
VOLUMETTO IV
\begin{itemize}
 { \item[{7.}]Fundamental characters of the group of permutations of $f$ objects }{}
 { \item[{14.}]Cubic symmetry }{}
 { \item[{25.}]Spherical functions with spin (s=1) }{}
 { \item[{29.}]Spherical functions with spin (II) }{} \\
\end{itemize}
VOLUMETTO V
\begin{itemize}
\item[{1.}]Representations of the Lorentz group
\item[{6.}]Spinor transformations
\item[{7.}]Spherical functions with spin (s=1/2)
\item[{8.}]Infinite-dimensional unitary representations of the Lorentz group
\end{itemize}
It is remarkable that the above list approximately mimics the
sequence of arguments covered by the Weyl book \cite{Weyl28}.
Explicit reference to this book appear in Sec. 21 of Volumetto
III; actually, the development of the subject of this section
follows closely that of Sec. 3, chapter 4 of \cite{Weyl28}
(Intensity and selection rules). Likely as previous sections, it
reiterates Weyl's arguments, but the last section of the notebooks
dealing with group-theoretical applications (last entry in the
above list) is a preliminary study of what will be one of the most
important (published) papers by Majorana on a generalization of
the Dirac equation to particles with arbitrary spin
\cite{infinite}. We will get an insight into this and other
subject in the following section.

${}$ \\

\section{\label{4} CONFRONTING MAJORANA WITH THE WEYL APPROACH}

Here we now compare the results achieved by the two authors on
some relevant specific topics of Group Theory and its applications
to Quantum mechanics.

${}$ \\

\subsection{Group Theory}

As a starting point let us consider the discussion of the group of
unitary transformations in two dimensions. Weyl approach is
typical of a mathematical physicist: he gives an abstract
description of the above-mentioned group as a subgroup of the
group of linear transformations:
\begin{equation} \label{4.1}
\begin{array}{rcl}
x \rightarrow x^\prime &=& \alpha x + \beta y , \\
y \rightarrow y^\prime &=& - \beta^\ast x + \alpha^\ast y ,
\end{array}
\end{equation}
with $\alpha \alpha^\ast + \beta \beta^\ast =1$. Since $\alpha$
and $\beta$ are, in general, complex numbers, Weyl characterizes
the transformations of the group by means of 4 real parameters
(real and imaginary parts of $\alpha$ and $\beta$) the sum of
whose squares is 1. By using this representation, he then points
out (without an explicit proof) that the composition of two such
transformations corresponds to the multiplication rule for
Hamilton quaternions.

Instead, in the Volumetti, Majorana gives a very detailed
description of the group considered, reporting a simple proof of
the above-mentioned relationship with the quaternions. He writes
the transformation in Eq. (\ref{4.1}) in matrix form:
\begin{equation}\label{4.2}
 \left( \begin{array}{c} x^\prime \\ y^\prime \end{array} \right)
\; = \; \left(\ba{cc} k \, + \, i \, \lambda & ~ - \, \mu \, + \, i \, \nu \\
  \mu \, + \, i \, \nu & ~ k \, - \, i \, \lambda
  \ea \right) \left( \begin{array}{c} x \\ y \end{array} \right)
  ,
\end{equation}
where $k,\lambda, \mu, \nu$ are 4 real parameters (in practice,
according to Weyl, the real and imaginary parts of the complex
numbers $\alpha$ and $\beta$ above). The composition of two
unitary transformations $T_A \cdot T_B$ is then described by
matrix multiplication $A B$:
\begin{equation}
\begin{array}{c}
\displaystyle A  =  \left(\ba{cc}
  k \, + \, i \, \lambda & ~ - \, \mu \, + \, i \, \nu \\
  \mu \, + \, i \, \nu & ~ k \, - \, i \, \lambda
  \ea \right) , \quad
B  =  \left(\ba{cc}
  k^\prime \, + \, i \, \lambda^\prime & ~ - \, \mu^\prime \, + \, i \,
  \nu^\prime \\
  \mu^\prime \, + \, i \, \nu^\prime & ~ k^\prime \, - \, i \,
  \lambda^\prime
  \ea \right) \nonumber , \\ {} \\
\displaystyle A \, B  =
 \left(\ba{cc}
  k^{\prime \prime} \, + \, i \, \lambda^{\prime \prime} & ~ - \, \mu^{\prime \prime}
  \, + \, i \, \nu^{\prime \prime} \\
  \mu^{\prime \prime} \, + \, i \, \nu^{\prime \prime} & ~ k^{\prime \prime} \, - \, i \,
  \lambda^{\prime \prime}
  \ea \right) ,
 \ea
 \ee
so that the correspondence with the multiplication rule for
quaternions immediately follows:
\begin{equation}\label{4.3}
  \ba{rcl}
k^{\prime \prime} & = & k k^\prime - \lambda \lambda^\prime - \mu
\mu^\prime - \nu \nu^\prime , \\
\lambda^{\prime \prime} & = & k \lambda^\prime + \lambda k^\prime
-
\mu \nu^\prime + \nu  \mu^\prime , \\
\mu^{\prime \prime} & = & k \mu^\prime + \lambda \nu^\prime + \mu
k^\prime - \nu \lambda^\prime , \\
\nu^{\prime \prime} & = & k \nu^\prime - \lambda \mu^\prime + \mu
\lambda^\prime + \nu k^\prime .
  \ea
\end{equation}
A general derivation of the generators of the group is reported in
the Volumetti as well. Moreover, keeping in mind physical
applications, Majorana also gives explicit expressions for matrix
representations corresponding to the index (angular momentum)
$j=0,1/2,1,3/2,2$ (even in modern texts, only $j=0$ and $j=1/2$
representations are reported).

Similar considerations hold for the study of both the group of
rotations and the Lorentz group.

Instead, for what concerns the group of permutations, relevant for
applications to a system of identical particles, Majorana
completes and generalizes the analysis by Weyl, by giving the
explicit expressions for the fundamental characters of the group.
He also discusses cubic symmetry, not considered in \cite{Weyl28},
by considering the group of permutations of 4 objects, and its
relationship with the 24 (proper) rotations transforming the
$x,y,z$ axes into themselves.

${}$ \\

\subsection{Applications to Quantum Mechanics}

\begin{figure}
\epsfysize=16cm \epsfxsize=12cm \centerline{\epsffile{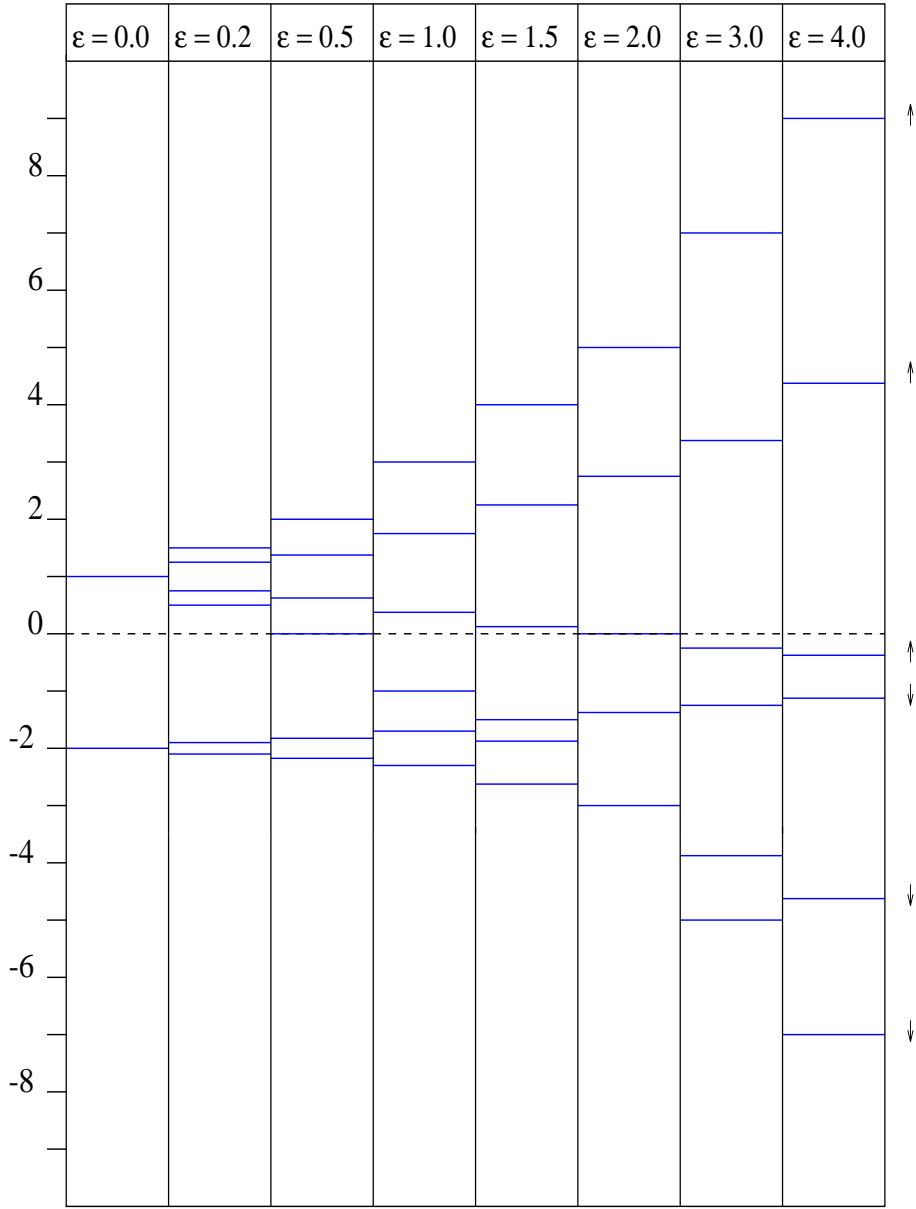}}
\caption{Transition from the anomalous Zeeman effect to the
Paschen-Back effect, according to Majorana \cite{volumetti}. The
parameter $\epsilon$ is the dimensionless Larmor frequency,
parameterizing the perturbation on the system induced by the
applied magnetic field. On the left axis the energy (in arbitrary
units) of the spectral lines is reported while, on the right, the
spin up or spin down component of the given line in the
Paschen-Back effect is indicated.} \label{zeeman}
\end{figure}

Several applications of Group Theory to quantum mechanical
phenomena are considered by Weyl as well as Majorana. We have
already pointed out in Sec. \ref{3} that the question of selection
rules for atomic transitions in a central field is tackled in the
same way by the two authors, with quite similar results. Here we
focus on another typical problem of atomic physics, namely that of
the anomalous Zeeman effect, that Weyl considers as a ``simple''
application of the theory of the group of rotations. He evaluates
the Land\'e $g$-factor (for one-electron as well as many-electron
atoms) entering in the anomalous splitting of the spectral lines
just as the result of the mean expectation values of some angular
momentum operators (adopting a ``physical language'', according to
Weyl himself), showing the power of the group-theoretical methods
(but even its uncomfortable use for physicists). On the contrary
Majorana deals with the problem as with an exercise of physics,
starting from the Dirac equation for the physical system
considered. The appropriate Hamiltonian matrix describing the
system is, then, deduced and the energy eigenvalues are obtained
with the help of Group Theory (following the Weyl method).
Majorana reports detailed predictions of the physical results for
the energy eigenvalues in the weak field limit as well as for
strong magnetic fields and, moreover, he considers the transition
from the anomalous Zeeman effect to the Paschen-Back effect by
changing the size of the magnetic field acting on the atom. His
results for the shift of the spectral lines are sketched in Figure
\ref{zeeman}. It appears, by looking at the notebooks, that such
transition is studied only on qualitative grounds; nevertheless
some details in Sec. 22 of Volumetto III seem to suggest some
specific calculations which, to our knowledge, cannot be performed
by using perturbation theory but only numerically.

${}$ \\

\subsection{Representations of the Lorentz Group}

In \cite{Weyl28} Weyl supposed that the general properties of
finite-dimensional groups are preserved in when passing to
infinite-dimensional ones. Majorana, instead, explored such a
connection in a different way, and this is particularly evident in
his treatment of the Lorentz group. This group underlies the
Theory of Relativity and its representations are especially
relevant for the Dirac equation in Relativistic Quantum Mechanics.
In the Weyl book only a particular kind of such representations
are considered (those related to the two-dimensional
representations of the group of rotations, according to Pauli),
but we must observe that an exhaustive study of this subject was
still lacking at that time. The correspondence between the Dirac
equation and the Lorentz transformations is obviously pointed out
but, surprisingly, the group properties of this connection are not
highlighted.

Majorana treats the Lorentz group in his third and fifth notebook.
Confining ourselves to the same topics covered also by Weyl, again
he gives a detailed deduction of the relationship between the
representations of the Lorentz group and the matrices of the
(special) unitary group in two dimensions, and a strict connection
with the Dirac equation is always taken into account. Moreover the
{\it explicit} form of the transformations of every bilinear in
the spinor field $\Psi$ is reported. For example, he obtains that
some of such bilinears behave as the 4-position vector
$(ct,x,y,z)$ or as the components of the rank-2 electromagnetic
tensor $({\bvec{E}},{\bvec{H}})$ under Lorentz transformations,
according to the following rules:
 \begin{eqnarray}
& \Psi^\dagger \Psi \sim - i \Psi^\dagger \alpha_x \alpha_y
\alpha_z \Psi \sim c t, &  \\
& - \Psi^\dagger \alpha_x \Psi \sim i \Psi^\dagger
\alpha_y \alpha_z \Psi \sim x, &  \\
& - \Psi^\dagger \alpha_y \Psi \sim i \Psi^\dagger \alpha_z
\alpha_x \Psi \sim y, &   \\
& - \Psi^\dagger \alpha_z \Psi \sim i
\Psi^\dagger \alpha_x \alpha_y \Psi \sim z, & \\
& i \Psi^\dagger  \beta \alpha_x \Psi  \sim  E_x,\quad
 i \Psi^\dagger  \beta \alpha_y \Psi  \sim  E_y, \quad
 i \Psi^\dagger  \beta \alpha_z \Psi  \sim  E_z,  &   \\
& i \Psi^\dagger  \beta \alpha_y \alpha_z \Psi \sim  H_x ,\quad
 i \Psi^\dagger  \beta \alpha_z \alpha_x \Psi \sim  H_y ,\quad
 i \Psi^\dagger  \beta \alpha_x \alpha_y \Psi \sim  H_z, &  \\
& \Psi^\dagger  \beta \Psi \; \sim \; \Psi^\dagger  \beta \alpha_x
\alpha_y \alpha_z \Psi \; \sim \; 1, &
 \end{eqnarray}
where $\alpha_x, \alpha_y, \alpha_z, \beta$ are Dirac matrices.

But, probably, the most important result achieved by Majorana on
this subject is his discussion of {\bf infinite-dimensional}
unitary representations of the Lorentz group, giving also an {\it
explicit} form for them. Note that such representations were
independently discovered by Wigner in 1939 and 1948
\cite{rediscover} and were thoroughly studied only in the years
1948-1958 \cite{Gelfand}. Lucky enough, we are able to reconstruct
the reasoning which led Majorana to discuss the
infinite-dimensional representations. In Sec. 8  of Volumetto V we
read:
\begin{quote}
``The representations of the Lorentz group are, except for the
identity representation, essentially not unitary, i.e., they
cannot be converted into unitary representations by some
transformation. The reason for this is that the Lorentz group is
an open group. However, in contrast to what happens for closed
groups, open groups may have irreducible representations (even
unitary) in infinite dimensions. In what follows, we shall give
two classes of such representations for the Lorentz group, each of
them composed of a continuous infinity of unitary
representations.''
\end{quote}
The two classes of representations correspond to integer and
half-integer values for the representation index $j$ (angular
momentum). Majorana begins by noting that the group of the real
Lorentz transformations acting on the variables $ct, x, y, z$ can
be constructed from the infinitesimal transformations associated
to the matrices (not taken into account by Weyl):
\begin{equation}
\begin{array}{c}
${}$ \\
\displaystyle S_x \; = \; \left( \ba{cccc}
                 0 & 0 & 0 & 0  \\
                 0 & 0 & 0 & 0  \\
                 0 & 0 & 0 & -1 \\
                 0 & 0 & 1 & 0
             \ea \right) ,\quad
S_y \; = \; \left(  \ba{cccc}
                 0 & 0 & 0 & 0 \\
                 0 & 0 & 0 & 1 \\
                 0 & 0 & 0 & 0 \\
                 0 & -1& 0 & 0
             \ea \right), \quad
S_z \; = \; \left(  \ba{cccc}
                 0 & 0 & 0 & 0 \\
                 0 & 0 &-1 & 0 \\
                 0 & 1 & 0 & 0 \\
                 0 & 0 & 0 & 0
             \ea \right),
\\ ${}$ \\ \displaystyle
T_x \; = \; \left( \ba{cccc}
                 0 & 1 & 0 & 0 \\
                 1 & 0 & 0 & 0 \\
                 0 & 0 & 0 & 0 \\
                 0 & 0 & 0 & 0
             \ea \right) ,\quad
T_y \; = \; \left(  \ba{cccc}
                 0 & 0 & 1 & 0 \\
                 0 & 0 & 0 & 0 \\
                 1 & 0 & 0 & 0 \\
                 0 & 0 & 0 & 0
             \ea \right) , \quad
T_z \; = \; \left(  \ba{cccc}
                 0 & 0 & 0 & 1 \\
                 0 & 0 & 0 & 0 \\
                 0 & 0 & 0 & 0 \\
                 1 & 0 & 0 & 0
             \ea \right),
\end{array}
 \end{equation}
from which he deduces the general commutation relations satisfied
by the $S$ and $T$ operators acting on generic (even infinite)
tensors or spinors:
 \bea
S_x \, S_y \, - \, S_y \, S_x &=&  S_z ,\nonumber \\ T_x \, T_y \,
- \, T_y \, T_x &=& - \, S_z, \nonumber \\ S_x \, T_x \, - \, T_x
\, S_x &=&
0, \label{4.3a} \\ S_x \, T_y \, - \, T_y \, S_x &=& T_z, \nonumber \\
S_x \, T_z \, - \, T_z \, S_x &=& - \, T_y, \nonumber \\
{\rm{etc}}. \nonumber
 \eea
Next he introduces the matrices
 \be \label{4.4}
a_x \; = \; i \, S_x,\quad b_x \; = \; - i \, T_x,\quad
{\rm{etc.}}
 \ee
which are Hermitian for unitary representations (and viceversa),
and obey the following commutation relations:
 \bea
\left[ a_x ,~ a_y \right] &=& i \, a_z, \nonumber \\
\left[ b_x ,~ b_y \right] &=& - i \, a_z, \nonumber \\
\left[ a_x ,~ b_x \right] &=& 0, \label{4.4a} \\
\left[ a_x ,~ b_y \right] &=& i \, b_z,  \nonumber \\
\left[ b_x ,~ a_y \right] &=& i \, b_z,  \nonumber \\
 {\rm{etc}}. \nonumber
 \eea
By using only these relations he then obtains (algebraically
\footnote{The algebraic method to obtain the matrix elements in
Eq. (\ref{4.4b}) follows closely the analogous one for evaluating
eigenvalues and normalization factors for angular momentum
operators, discovered by Born, Heisenberg and Jordan in 1926 and
reported in every textbook on Quantum Mechanics (see, for example,
\cite{Sakurai}).}) the explicit expressions of the matrix elements
for given $j$ and $m$ \cite{volumetti} \cite{infinite}. The
non-zero elements of the infinite matrices $a$ and $b$, whose
diagonal elements are labelled by $j$ and $m$, are as follows:
 \bea
% \ba{rcl}
\!\!\!\!\! \dps
<j,m \, | \, a_x - i a_y \, | \, j,m+1> &=& \dps \sqrt{(j+m+1)(j-m)}, \nonumber \\
%& & \\
\!\!\!\!\! \dps
<j,m \, | \, a_x + i a_y \, | \, j,m-1> &=& \dps \sqrt{(j+m)(j-m+1)}, \nonumber \\
%& & \\
\!\!\!\!\! \dps
<j,m \, | \, a_z \, | \, j,m> &=& \dps m, \nonumber \\
%& & \\
\!\!\!\!\! \dps <j,m \, | \, b_x - i b_y \, | \, j+1,m+1> &=& \dps
- \, \frac{1}{2} \,
\sqrt{(j+m+1)(j+m+2)}, \nonumber \\
%& & \\
\!\!\!\!\! \dps <j,m \, | \, b_x - i b_y \, | \, j-1,m+1> &=& \dps
\frac{1}{2} \,
\sqrt{(j-m)(j-m-1)}, \label{4.4b} \\
%& & \\
\!\!\!\!\! \dps <j,m \, | \, b_x + i b_y \, | \, j+1,m-1> &=& \dps
\frac{1}{2} \,
\sqrt{(j-m+1)(j-m+2)}, \nonumber \\
%& & \\
\!\!\!\!\! \dps <j,m \, | \, b_x + i b_y \, | \, j-1,m-1> &=& \dps
- \, \frac{1}{2} \,
\sqrt{(j+m)(j+m-1)},  \nonumber \\
%& & \\
\!\!\!\!\! \dps <j,m \, | \, b_z \, | \, j+1,m> &=& \dps
\frac{1}{2} \, \sqrt{(j+m+1)(j-m+1)}, \nonumber \\
%& & \\
\!\!\!\!\! \dps <j,m \, | \, b_z \, | \, j-1,m> &=& \dps
\frac{1}{2} \, \sqrt{(j+m)(j-m)}. \nonumber
% \ea
 \eea
The quantities on which $a$ and $b$ act are infinite tensors or
spinors (for integer or half-integer $j$, respectively) in the
given representation, so that Majorana effectively constructs, for
the first time, infinite-dimensional representations of the
Lorentz group. In \cite{infinite} the author also picks out a
physical realization for the matrices $a$ and $b$ for Dirac
particles with energy operator $H$, momentum operator $\bvec{p}$
and spin operator $\bvec{\sigma}$:
\begin{equation} \label{4.5}
 a_x \, = \, \frac{1}{\hbar} \left( y p_z - z p_y \right) +
 \frac{1}{2} \sigma_x , \quad b_x \, = \, \frac{1}{\hbar} x
 \frac{H}{c} + \frac{i}{2} \alpha_x , \quad {\rm{etc.}} ,
\end{equation}
where $\alpha_x, \alpha_y,\alpha_z$ are the Dirac
$\alpha$-matrices.

Further development of this material then brought Majorana to
obtain a relativistic equation for a wave-function $\psi$ with
infinite components, able to describe particles with arbitrary
spin (the result was published in 1932 \cite{infinite}). By
starting from the following variational principle:
\begin{equation}\label{4.5a}
  \delta \int \ov{\psi} \left( \frac{H}{c} + {\bvec{\alpha}} \cdot
  {\bvec{p}} - \beta m c \right) \psi \, d^4 x \, = \, 0 ,
\end{equation}
By requiring the relativistic invariance of the variational
principle in Eq. (\ref{4.5a}), Majorana deduces both the
transformation law for $\psi$ under an (infinitesimal) Lorentz
transformation and the explicit expressions for the matrices
${\bvec{\alpha}}$, $\beta$. In particular, the transformation law
for $\psi$ is obtained directly from the corresponding ones for
the variables $ct, x, y, z$ by means of the matrices $a$ and $b$
in the representation (\ref{4.5}). By using the same procedure
leading to the matrix elements in (\ref{4.4b}), Majorana gets the
following expressions for the elements of the (infinite) Dirac
$\bvec{\alpha}$ and $\beta$ matrices:
 \bea
\!\!\!\!\! \dps <j,m \, | \, \alpha_x - i \alpha_y \, | \,
j+1,m+1> &=& \dps - \,  1/2  \, \sqrt{\frac{(j+m+1)(j+m+2)}{\dps
\left( j+ 1/2  \right) \left( j +  3/2  \right)}},
\nonumber \\
%& & \\
\!\!\!\!\! \dps <j,m \, | \, \alpha_x - i \alpha_y \, | \,
j-1,m+1> &=& \dps - \,  1/2  \, \sqrt{\frac{(j-m)(j-m-1)}{\dps
\left( j- 1/2  \right) \left( j +  1/2  \right)}},
\nonumber \\
%& & \\
\!\!\!\!\! \dps <j,m \, | \, \alpha_x + i \alpha_y \, | \,
j+1,m-1> &=& \dps  1/2  \, \sqrt{\frac{(j-m+1)(j-m+2)}{\dps \left(
j+ 1/2  \right) \left( j +  3/2  \right)}},
\nonumber \\
%& & \\
\!\!\!\!\! \dps <j,m \, | \, \alpha_x + i \alpha_y \, | \,
j-1,m-1> &=& \dps   1/2  \, \sqrt{\frac{(j+m)(j+m-1)}{\dps \left(
j- 1/2  \right) \left( j +  1/2  \right)}},
\label{4.5b}  \\
%& & \\
\!\!\!\!\! \dps <j,m \, | \, \alpha_z \, | \, j+1,m> &=& \dps - \,
 1/2  \, \sqrt{\frac{(j+m+1)(j-m+1)}{\dps \left(
j+ 1/2  \right) \left( j +  3/2  \right)}},
\nonumber \\
%& & \\
\!\!\!\!\! \dps <j,m \, | \, \alpha_z \, | \, j-1,m> &=& \dps - \,
 1/2  \, \sqrt{\frac{(j+m)(j-m)}{\dps \left( j- 1/2
\right) \left( j +  1/2  \right)}}, \nonumber \\
%& & \\
\!\!\!\!\! \dps \beta &=& \frac{1}{\dps j +  1/2 } . \nonumber
% \ea
 \eea
The Majorana equation for particles with arbitrary spin has, then,
the same form of the Dirac equation:
\begin{equation}\label{4.6}
  \left( \frac{H}{c} + {\bvec{\alpha}}\cdot {\bvec{p}} - \beta m c
  \right) \psi \, = \, 0 ,
\end{equation}
but with different (and infinite) matrices $\alpha$ and $\beta$,
whose elements are given in Eqs. (\ref{4.5b}). The rest energy of
the particles thus described has the form:
\begin{equation}\label{4.7}
  E_0 \, = \, \frac{m c^2}{\displaystyle s + 1/2} ,
\end{equation}
and depends on the spin $s$ of the particle. We here stress that
the scientific community of that time was convinced that only
equations of motion for spin 0 (Klein-Gordon equation) and spin
1/2 (Dirac equation) particles could be written down. The
importance of the Majorana work was first realized by van der
Waerden \cite{vdw} but, unfortunately, the paper remained
unnoticed until recent times.

${}$ \\

\section{\label{5} FINAL RESULTS}

As emerged from the above, when Majorana became aware of the great
relevance of the Weyl's application of the Group Theory to Quantum
Mechanics, he immediately grabbed the Weyl method and developed it
in many applications. What outlined in the previous sections gives
account of only a partial look of the question involving the
adoption and application of the new method, that arose just after
the appearance of the Weyl book. There is no trace of further work
in this direction by Weyl after 1928-1931, as discussed in Sec.
\ref{2}, while Majorana continued to use Group Theory in his
scientific work (see Secs. \ref{3} and \ref{4}) till up 1938 when
he lectured on this subject. Indeed the Weyl proposal for a
``novel'' description of Quantum Mechanics was completely adopted
by Majorana, who applied it in his lectures on Theoretical Physics
at the University of Naples (January - March 1938). In these
lectures \cite{lezioni} he followed Weyl in ``sandwiching''
mathematical formalism with physics applications but, differently
from the author of {\it Gruppentheorie und Quantenmechanik}, he
started from physics rather than mathematics, and the physical
viewpoint was underlying throughout the lectures.

We finish by reporting some passages from the starting lecture
held in Naples by Majorana, revealing an attitude which, some
decades after the appearance of the Weyl proposal, became the
usual one for authors dealing with Quantum Mechanics:
\begin{quote}
`` In order to illustrate Quantum Mechanics in its present form,
two nearly opposite methods could apply. The first one is the
so-called historical method..., the other is the mathematical
one... Both methods, if applied in an exclusive way, present very
serious problems...

Then, if we simply illustrate the theory according to its
historical appearance, we at first would make the listener feel
ill-at-ease or create suspicion... that nowadays is no longer
justified. The second method... does not at all allow to
understand the genesis of the formalism... and, above all, it
completely disappoints the desire of guessing its physical meaning
in some manner.

The only way which can be followed..., without leaving off
anything of the historical genesis of the ideas or the language
itself which at the moment dominate, is to put before an as wide
and clear as possible exposition of the essential mathematical
tools...

Our only ambition will be to illustrate as clearly as possible the
effective use of these tools made by physicist in over a decade,
in which use - that never led to difficulties or ambiguities - is
the source of their certainty.''
\end{quote}

\vspace{2cm} \noindent {\Large \bf ACKNOWLEDGMENTS}

% If you have acknowledgments, this puts in the proper section head.
%\begin{acknowledgments}

\noindent We are indebted with Prof. E. Recami, Prof. A. Sciarrino
and Dr. E. Majorana jr for many fruitful discussions. This work
has been partially supported by COFIN funds (coordinated by P.
Tucci) of the Italian MIUR.

%\end{acknowledgments}

% Create the reference section using BibTeX:
%\bibliography{basename of .bib file}

%\vspace{2cm}

\end{document}